\documentclass[onecolumn,journal,11pt]{IEEEtran}
\usepackage[cmex10]{amsmath}
\usepackage{%
amssymb%
,mathrsfs%
,amsthm
}
\usepackage[noadjust]{cite}
\usepackage{array,url}
\usepackage{cases}
\usepackage{ytableau,tikz,varwidth}
\usepackage{here}

\usepackage[latin1]{inputenc}
\usepackage[english]{babel} 

\newtheorem{theorem}{Theorem}{\bfseries}{\it}
\newtheorem{definition}{Definition}{\bfseries}{\it}
\newtheorem{proposition}{Proposition}{\bfseries}{\itshape}
\newtheorem{lemma}{Lemma}{\bfseries}{\itshape}
\newtheorem{corollary}{Corollary}{\bfseries}{\itshape}
{\bfseries}{\itshape}
{\bfseries}{\itshape}
{\bfseries}{\itshape}
{\bfseries}{\itshape}
{\bfseries}{\itshape}
\newtheorem{remark}{Remark}{\bfseries}{\itshape}
\newtheorem{exemple}{Exemple}{\bfseries}{\itshape}
\newtheorem{notation}{Notation}{\bfseries}{\itshape}

\renewcommand{\leq}{\leqslant} 
\renewcommand{\le}{\leqslant} 
\renewcommand{\geq}{\geqslant} 
\renewcommand{\ge}{\geqslant}

\begin{document}

\title{Algebraic Properties of Polar Codes From a New Polynomial Formalism}

\author{\IEEEauthorblockN{Magali Bardet\IEEEauthorrefmark{1}, Vlad Dragoi\IEEEauthorrefmark{1}, Ayoub Otmani\IEEEauthorrefmark{1},  Jean-Pierre Tillich\IEEEauthorrefmark{2}\\}
\IEEEauthorblockA{\IEEEauthorrefmark{1}Normandie Univ, France; UR, LITIS, F-76821 Mont-Saint-Aignan, France\\
\texttt{\{magali.bardet,vlad.dragoi,ayoub.otmani\}@univ-rouen.fr}\\}
\IEEEauthorblockA{\IEEEauthorrefmark{2}Inria, SECRET Project, 78153 Le Chesnay Cedex, France\\
\texttt{\{jean-pierre.tillich\}@inria.fr}}
}

\maketitle

\begin{abstract}
Polar codes form a very powerful family of codes  with a low complexity decoding algorithm that  attains 
many information theoretic limits in error correction and source coding.  These codes are closely related
to Reed-Muller codes because both can be described with the same algebraic formalism, namely 
they  are generated by evaluations of monomials.
However, finding the right set of generating monomials for a polar code which optimises the decoding performances is a nontrivial task and is channel dependent. The  purpose of this paper 
is to reveal some universal properties of these monomials. 
We will namely prove that there is a way to define a nontrivial (partial) order on monomials 
so that the monomials generating a polar code devised for a binary-input symmetric channel
always form a decreasing set. We call such codes decreasing monomial codes.
 The fact that polar codes are decreasing monomial codes turns out to have rather deep consequences on their structure. 
 Indeed, we show that decreasing monomial codes have a very large permutation group by proving that it contains a group called lower triangular  affine group. Furthermore, the codewords of minimum weight  correspond exactly to the orbits of the minimum weight codewords  that are obtained from evaluations
of monomials of the generating set. In particular, it gives an efficient way of counting the number of minimum weight codewords of a decreasing monomial code and henceforth of a polar code. 
\end{abstract}

\begin{IEEEkeywords}
Polar codes, Reed-Muller codes, capacity-achieving codes, decreasing monomial codes, lower triangular  affine group, post-quantum cryptography.
\end{IEEEkeywords}

\newcommand{\card}[1]{|#1|}

\renewcommand{\epsilon}{\varepsilon}
\renewcommand{\phi}{\varphi}
\renewcommand{\leq}{\leqslant}
\renewcommand{\le}{\leqslant}
\renewcommand{\geq}{\geqslant}
\renewcommand{\ge}{\geqslant}

\newcommand{\dec}{\mathcal{D}}

\newcommand{\eqdef}{\stackrel{\text{def}}{=}}

\newcommand{\A}{\mathbb{A}}
\newcommand{\C}{\mathbb{C}}
\newcommand{\F}{\mathbb{F}}
\newcommand{\K}{\mathbb{K}}
\newcommand{\N}{\mathbb{N}}
\newcommand{\R}{\mathbb{R}}
\newcommand{\T}{\mathbb{T}}
\newcommand{\Z}{\mathbb{Z}}

\newcommand{\Fc}{\mathcal{F}}
\newcommand{\Gc}{\mathcal{G}}
\newcommand{\Ic}{\mathcal{I}}
\newcommand{\Jc}{\mathcal{J}}
\newcommand{\Xc}{\mathcal{X}}
\newcommand{\Yc}{\mathcal{Y}}
\newcommand{\Zc}{\mathcal{Z}}

\newcommand{\Rm}{{\mathbf {R}}_m}
\newcommand{\Rmp}{{\mathbf {R}}_{m+1}}
\newcommand{\Pe}{\left[\mathcal{P}\right]}
\newcommand{\he}[1]{\left[h_{#1}\right]}
\newcommand{\FXT}{\mathbb F_2[X]/(X^p-1)/\cong}
\newcommand{\Fx}{\mathbb F_2[x]/(x^p-1)}
\newcommand{\Zp}{\mathbb Z_p^*}
\newcommand{\ord}{\ensuremath{\mathsf{ord}}}
\newcommand{\supp}{\mathsf{supp}}
\newcommand{\suppn}[1]{\overline{\mathsf{supp}({#1})}}
\newcommand{\eea}{\ensuremath{\textsf{EEA}}}
\newcommand{\wt}{\mathsf{wt}}

\newcommand{\fq}{\F_q}
\newcommand{\ft}{\F_2}
\newcommand{\ftm}{\F_{2^m}}
\newcommand{\fqm}{\F_{q^m}}

\newcommand{\word}[1]{\ensuremath{\boldsymbol{#1}}}
\newcommand{\mat}[1]{\boldsymbol{#1}}
\newcommand{\code}[1]{\mathscr{#1}}
\newcommand{\CC}{\code{C}}
\newcommand{\DD}{\code{D}}
\newcommand{\RM}{\code{R}}

\newcommand{\av}{\word{a}}
\newcommand{\bv}{\word{b}}
\newcommand{\cv}{\word{c}}
\newcommand{\dv}{\word{d}}
\newcommand{\ev}{\word{e}}
\newcommand{\gv}{\word{g}}
\newcommand{\hv}{\word{h}}
\newcommand{\mv}{\word{m}}
\newcommand{\sv}{\word{s}}
\newcommand{\uv}{\word{u}}
\newcommand{\vv}{\word{v}}
\newcommand{\wv}{\word{w}}
\newcommand{\xv}{\word{x}}
\newcommand{\yv}{\word{y}}
\newcommand{\zv}{\word{z}}

\newcommand{\Gv}{\word{G}}
\newcommand{\Bv}{\word{B}}
\newcommand{\Qv}{\word{Q}}
\newcommand{\Sv}{\word{S}}
\newcommand{\Uv}{\word{U}}
\newcommand{\Vv}{\word{V}}
\newcommand{\Wv}{\word{W}}
\newcommand{\Zv}{\word{Z}}

\newcommand{\Am}{\mat{A}}
\newcommand{\Bm}{\mat{B}}
\newcommand{\Dm}{\mat{D}}
\newcommand{\Gm}{\mat{G}}
\newcommand{\Gp}{\mat{G}}
\newcommand{\Gs}{\mat{G}_s}
\newcommand{\Hm}{\mat{H}}
\renewcommand{\Im}{\mat{I}}
\newcommand{\Mm}{\mat{M}}
\newcommand{\Nm}{\mat{N}}
\newcommand{\Om}{\mat{0}}
\newcommand{\Pm}{\mat{P}}
\newcommand{\Sm}{\mat{S}}
\newcommand{\Vm}{\mat{V}}

\newcommand{\Mon}{\mathcal{M}_m}
\newcommand{\eval}{\mathsf{ev}}
\newcommand{\rmo}{{r_{-}}}
\newcommand{\rpl}{{r_{+}}}
\newcommand{\Wm}{W_{\text{min}}}
\newcommand{\Alow}{{\rm LTA}(m,2)}
\newcommand{\orbit}[1]{\mathcal O_{#1}}
\newcommand{\Aff}{{\mathbb A}}
\newcommand{\GL}{{\rm GL}}

\newcommand{\cwp}{\star} 
\newcommand{\sqc}[1]{{#1}^2} 
\newcommand{\sqcp}[1]{\left(#1\right)^2}  
\newcommand{\stp}[2]{#1\cwp#2}   
\newcommand{\scp}[2]{#1\cdot #2} 
\newcommand{\sh}[2]{{\mathcal{S}}_{#1}\left(#2\right)}
\newcommand{\pu}[2]{{\mathcal{P}}_{#1}\left(#2\right)}
\newcommand{\Iind}{{\mathcal{I}}}
\newcommand{\Jind}{{\mathcal J}}

\newcommand{\indices}{\text{\rm ind}}

\newcommand{\cmin}{\cv_{\text{min}}}
\newcommand{\dmaxf}{{d_{f}^{\text{max}}}}
\newcommand{\dmaxm}{{d_{m}^{\text{max}}}}

\newcommand{\plus}[1]{{#1^+}}
\newcommand{\plusb}[1]{{\left(#1\right)^+}}
\newcommand{\moins}[1]{{#1^-}}
\newcommand{\moinsb}[1]{{\left(#1\right)^-}}
\newcommand{\Bha}[1]{{\mathcal{B}\left( #1 \right)}}
\newcommand{\W}[2]{W^{#1}_{#2}}
\newcommand{\Wb}[2]{\left(W^{#1}_{#2}\right)}
\newcommand{\Wg}[3]{{{#1}^{#2}_{#3}}}
\newcommand{\Wgb}[3]{\left({{#1}^{#2}_{#3}}\right)}
\newcommand{\f}[2]{f_{[#1..#2]}}

\newcommand{\Mo}{\mathcal{M}_m}
\newcommand{\Mop}{\mathcal{M}_{m+1}}

\newcommand{\weako}{\preceq_{\text{w}}}
\newcommand{\degrad}{\preceq_{\text{d}}}
\newcommand{\prob}{\mathbf{prob}}

\section{Introduction} \label{sec:intro}

{\bf Polar codes and Reed Muller codes viewed as monomial codes.} 
Polar codes were discovered by Ar{\i}kan \cite{A09} and form a very powerful family of codes that gave a nice constructive way of attaining many information theoretic limits in error correction and source coding. In particular, they allow to attain the capacity of any symmetric memoryless channel with a low complexity decoding algorithm (namely the successive cancellation decoder of Ar{\i}kan).
 These codes are closely related to Reed-Muller codes in the sense that they can both be described with the same algebraic formalism, namely as \emph{monomial} codes. 
 Monomial codes are evaluation codes where  a specific set of monomials provides a generator matrix. 
 A Reed-Muller code $\RM(r,m)$ is generated by the evaluation over $\F_2^m$ of \emph{all} monomials degree at most $r$
 in $m$ variables. 
%
A polar code of length $2^m$ is also generated by 
evaluation of monomials, but not necessarily by the same monomials as a Reed-Muller code: 
if we want a polar code of a certain dimension for a certain channel, we are going to take a very specific set of monomials
which is in general significantly different from the Reed-Muller choice. This choice will give good performances for the Ar{\i}kan 
successive cancellation decoder. It turns out  that this decoder is very closely related  to Dumer's recursive algorithm
 for decoding Reed-Muller codes \cite{D06a} based on the $(u|u+v)$ decomposition. Basically 
 Dumer's decoding algorithm is the successive cancellation decoder of Ar{\i}kan but the performance of the decoder is much worse in this case because
 the choice of monomials for a Reed-Muller code is not well suited to this kind of decoding algorithm. 
 
{\bf Polar codes are decreasing monomial codes.} Finding the right set of generating monomials for a polar code which optimises the decoding performances under the successive cancellation decoder is by no means an easy task (see for instance \cite{TV13}) and moreover it is channel
 dependent. Our purpose is here to reveal some universal properties of these monomials, where by ``universal'' we mean
 properties that do not depend on the channel. We will namely prove that, regardless of the binary-input symmetric channel the polar code is devised for, there is a way to define a 
 nontrivial partial order on monomials for which 
 a polar code is always generated by
 a \emph{decreasing set}, that is to say: if 
 a monomial lies 
 in the generating set 
 then all monomials that are smaller also belong to it. %
 This property turns out to have rather deep consequences on the structure of the polar code.
 We call {\em decreasing monomial code} a monomial code whose generating set of monomials forms a decreasing set. 
We will namely prove that such codes have some interesting properties: 
\begin{itemize} 
	\item they have a very large permutation group. \label{nice1} 
	\item the codewords of minimum weight 
	correspond to orbits of the minimum weight codewords  
	obtained from evaluations of monomials of the generating set.  \label{nice2}
\end{itemize}
In particular, we will 
give an efficient way of counting the number of minimum weight codewords of a decreasing monomial code and henceforth of a polar code. 

{\bf The permutation group of decreasing monomial codes.} 
The \emph{permutation group} of a code is the group of  the permutations of coordinates leaving the code globally invariant, \textit{i.e.} it permutes the coordinates of any codeword into another codeword. 
It is well known that the permutation group of a non-trivial Reed-Muller 
code $\RM(r,m)$ is isomorphic the whole affine group $\Aff_{m}$ over $\F_2^m$. This group is of size 
$2^{\Theta(m^2)}$  which is superpolynomial in the length $n=2^m$ of the Reed-Muller code, since it is of size $n^{\Theta(\log n)}$. 
It is also $2$-transitive and this property 
 has  been used recently to prove that Reed-Muller codes attain the capacity of the erasure channel \cite{KMSU15,KP15}.
 The fact that the size of the permutation group of a Reed-Muller code $\RM(r,m)$ is so large is related to the special choice of generating monomials of the code:
 the affine group actually acts in a natural way on monomials and transforms a monomial in the generating set into a sum of  monomials 
 of the generating set, since by an affine change of variables a monomial of degree less than or equal to $r$ is transformed into 
 a polynomial of degree less than or equal to $r$. 
 
 We do not expect such a behavior for polar codes, since the monomial generating set
  of the polar code has no reason to have the same property. However it will turn out that because
 of the fact that the set of monomials of the polar code is decreasing with our order 
 this set of monomials is transformed by the lower triangular affine group (corresponding to affine transformations $\xv \mapsto \Am \xv + \bv$ where
 $\Am$ is a lower triangular matrix with $1$'s on its diagonal) into a sum of monomials that still belong to the
 generating set. This will imply that the permutation group of a polar code, and of a decreasing monomial code in general, 
 contains a subgroup which is isomorphic to the lower triangular affine group. For a decreasing monomial code of length $2^m$ this subgroup is also of size $2^{\Theta(m^2)}$  which is also superpolynomial in the length $n=2^m$ of the code. In other words, 
 in a rather unexpected way, as in the case of Reed-Muller code the permutation group of a polar code is also extremely large (although it may only be one-transitive in this case).
 
 {\bf The structure of codewords of minimum weight in a decreasing monomial code.} The fact that the permutation group of a decreasing monomial code, and of a polar code in particular, is so large can be used for a better understanding of the structure of such codes.
 In particular we might expect to classify such codes as it has been done for affine invariant codes \cite{C90}. Here we are going to use
it to give a very convenient description of the minimal codewords. Indeed, a codeword of minimum weight is transformed into another minimal 
codeword by the action of the permutation group of the code. It turns out that this number of orbits is very small, since we are going to show that any such orbit contains a generating monomial of maximum degree. Therefore the number of such orbits is really small since it is at most of size $O(n)$ where $n$ is the length of the code. Moreover it is also rather easy to count the number of elements in the orbit and this will allow to count the number of codewords of minimum weight.

{\bf Other properties of decreasing monomial codes.} Decreasing monomial codes do not only have a large permutation group and
a convenient formula for counting the number of minimum weight codewords, they also display other nice features. For instance,
we will show that the dual of a decreasing monomial code is still a decreasing monomial code. It will also turn out that under a very mild and simple condition, a decreasing monomial code is weakly self-dual, meaning that it is contained in its dual. Polar codes
of rate sufficiently smaller than $\frac{1}{2}$ tend to meet this property as we will explain in what follows. This property 
has been used to devise quantum polar codes, see \cite{RDR12}. This family of codes is also closed under the star product of 
codes \cite{P92,R15} which can be used in decoding algorithms \cite{P92,K92}, for secure multi-party computation \cite{CCCX09,CCX11} or for cryptanalysis \cite{W10,FGOPT13,CGGOT14,COT14,CB14,CMP14}. 

All these properties of decreasing monomial codes shed some light on the structure of polar codes. They explain why the permutation of such codes is so large and give a convenient way of counting the minimum weight codewords. Both properties
turned to be essential to attack \cite{BCDOT16} the McEliece system based on polar codes \cite{SK14}. It might be interesting 
to use the fact that the permutation group is so large in order to get better decoding performances for the successive cancellation
decoder. The structure of the minimum weight codewords could also be used to devise rather tight bounds for maximum likelihood decoding of
polar codes. Finally we wish also to point out that this larger class of decreasing monomial codes might be interesting 
in itself. It contains both the polar code family and Reed-Muller codes. It might be interesting to study whether this 
class of codes attains the capacity of the erasure channel in particular (as has been done for Reed-Muller codes recently in 
\cite{KMSU15,KP15}). The fact that they have a very large permutation group might be a good hint that this might be possible
under very mild additional conditions with tools inspired by the \cite{KMSU15,KP15,TZ00} approach. 

\section{Reed-Muller, Monomial and Polar Codes} \label{sec:polar}
In this section we briefly review Reed-Muller codes, polar codes and the algebraic formalism we will use to describe both families.

\par{\bf Reed-Muller codes.}
It is well known that Reed-Muller codes of length $2^m$ can be obtained as evaluation codes of 
polynomials in $\F_2[x_0,\dots{},x_{m-1}]$. Polar codes can also be described through this formalism.
Since we are interested in evaluations of such polynomials over entries in $\F_2^m$ we 
will identify $x_i$ with $x_i^2$ and work in the ring $\Rm=\F_2[x_0,\dots{},x_{m-1}]/(x_0^2-x_0,\dots,x_{m-1}^2-x_{m-1})$. 
It will be convenient with this formalism to associate to a polynomial $g \in \Rm$ the binary vector denoted by 
$\eval(g)$ in
$\F_2^n$ with $n=2^m$ which is the evaluation of the polynomial in all the binary entries 
$\uv = (u_0,\dots,u_{m-1}) \in \F_2^m$. In other words 
$$
\eval(g) = \big(g(\uv) \big)_{\uv \in \F_2^m}
$$
With this notation, we view the indices 
as elements
of $\F_2^m$.  This notation does not specify the order we use for the
elements of $\F_2^m$.  We actually use the natural order by viewing
$\uv = (u_0,\dots,u_m)$ as the integer $\sum_{i=0}^{m-1}u_i 2^i$ where
$u_i\in\{0,1\}$. With this notation at hand, the Reed-Muller code
$\RM(r,m)$ is defined as
$$
\RM(r,m) \eqdef \big \{\eval(P) ~|~ P \in\Rm, \deg P \leq r \big\}
$$
The function $\eval: \Rm\to \F_2^n$ is an homomorphism of algebra.
Hence, the code $\RM(r,m)$ is generated by the codewords $\eval(g)$
where $g$ is a monomial of degree less than or equal to $r$. Recall
that a \emph{monomial} is any product of variables of the form
$x_0^{g_0} \cdots{} x_{m-1}^{g_{m-1}}$ where $g_0,\dots{},g_{m-1}$ are
binary.  The set of all monomials is denoted by:
$$
\Mon \eqdef \left \{x_0^{g_0} \cdots{} x_{m-1}^{g_{m-1}} ~|~ (g_0,\dots{},g_{m-1})\in\F_2^m \right \}.
$$ 
Reed-Muller codes have a very large permutation group which is isomorphic to the 
affine group over $\F_2^m$. Indeed, it can be checked that:
\begin{enumerate}
\item any bijective affine transformation $A$ over 
$\F_2^m$ can be viewed as a permutation of the code positions 
by mapping $(u_0,\dots,u_{m-1})$ to 
$A(u_0,\dots,u_{m-1})$;
\item this permutation 
leaves the code invariant since
$P(A(x_0,\dots,x_{m-1}))$ is a polynomial of degree at most the degree of $P$ and therefore if 
$\eval(P) \in \RM(r,m)$ then 
$\eval(P \circ A) \in \RM(r,m)$.
\end{enumerate}

\bigskip

\par{\bf Monomial codes.} Monomial codes form a very general family of codes that generalizes Reed-Muller codes. %
\begin{definition}[Monomial code]
Let $I\subseteq\Mon$ be a finite set of monomials in $m$ variables and set $n \eqdef 2^m$.
The linear code  defined by $I$ is the vector subspace $\CC(I) \subseteq \F_2^n$ 
generated by $\{ \eval(f) ~|~ f \in I\}$.
\end{definition}

The dimension of such codes is given by
\begin{proposition}\label{prop:dim_monomial}
For all $I \subseteq \Mon$ the dimension of the monomial code $\CC(I)$  is equal to $\card{I}$.
\end{proposition}
\begin{IEEEproof}
This comes from the linear independence of the monomials in $\Rm$ and
the fact that $\eval$ is an injective mapping from $\Rm$ to $\F_2^{2^m}$.
\end{IEEEproof}

\bigskip

\par{\bf Polar codes.} 
Recall that the Kronecker product of two matrices 
$\Am$ of size $r_a\times c_a$
and $\Bm = (b_{i,j})_{\begin{substack} 1 \le i \leq r_b, 1\le j \le
  c_b \end{substack}}$ 
defined over a same field is the $r_ar_b\times c_ac_b$
matrix  defined by: 
\[
\Am\otimes \Bm =\left(
\begin{array}{cccc}
  a_{1,1}\Bm & \cdots & a_{1,c_a}\Bm \\
  \vdots & b_{i,j} \Bm & \vdots \\
  a_{r_a,1}\Bm & \cdots & a_{r_a,c_a}\Bm\\
\end{array}
\right).
\]
%
What we call here a polar code is a binary polar code as defined by Ar{\i}kan in \cite{A09}.
 They can be described as codes
of length $n=2^m$, where $m$ is an arbitrary integer. They may take any dimension between 
$1$ 
and $2^m$. The polar code of length $n=2^m$ and dimension $k$ is obtained through 
a generator matrix which picks a specific subset of $k$ rows of the 
$ 2^m \times 2^m$ matrix: 
\[
\Gm_m \eqdef \underbrace{\begin{pmatrix} 1 & 1 \\ 0 & 1 \end{pmatrix} \otimes \cdots 
\otimes \begin{pmatrix} 1 & 1 \\ 0 & 1 \end{pmatrix}}_{m \;\text{ times}}.
\]
Note that we depart here slightly from the usual convention for polar codes which is
to use in the Kronecker product the matrix $\begin{pmatrix} 1 & 0 \\ 1 & 1 \end{pmatrix}$. The two
definitions (ours and the standard one) are easily seen to be equivalent, they just amount to order the
code positions differently. Our convention presents the advantage of simplifying the polynomial formalism
that follows. It is clear that a polar code is a monomial code. This comes from the fact that the rows of $\Gm_m$ are all possible evaluations of monomials. This fact is  proved by induction on $m$ by observing that $(1,1)$ is the evaluation over $\F_2$ of the constant monomial $1$ and that $(0,1)$ is the evaluation over $\F_2$ of the monomial $x_0$. If we consider the binary expansion of each row number (starting from $0$ to $2^m-1$) of $\Gm_m$ over $m$ bits 
 $i=\sum_{j=0}^{m-1}i_{j}2^j$ (where $i_j\in\{0,1\}$), 
then  the row of index $i$ of $\Gm_m$ is given by $\Gm_m[i]=\eval(x_0^{i_0}\dots x_{m-1}^{i_{m-1}}).$

The specific rows that are picked (or equivalently the choice of the generating monomials when we view a polar code as a monomial code) depends (a little bit) on the
noisy channel for which the code is devised. For a given noise model,
there is a way to compute the $k$ rows 
that defines the generator matrix.
We will use here again Ar{\i}kan's definition with the Bhatacharryya parameter. There are other definitions using 
either mutual information or the bit error probability. It turns out that whatever definition we use (Bhatacharryya parameter,
mutual information, bit error probability) polar codes will still be 
decreasing monomial codes and this will be essential here. 
The reason behind this is the use of a channel degradation argument 
to prove such a fact (see for instance Proposition \ref{prop:decreasing_degrading} or Lemma \ref{lem:almost_order}). All three parameters are monotone with respect to channel degradation
since Lemma \ref{lem:Bha_degradation} also holds when the Bhatacharryya parameter is replaced by mutual information or  the bit error probability.

In essence, constructing a polar code of dimension $k$ is equivalent to finding the $k$ ``best'' bit-channels that modelize the
channel that the decoder sees when it recovers one by one the information bits corresponding to the received codeword by the successive cancellation
decoder. We refer to \cite{A09} for the definition of the successive decoder and just give here the decision rule for choosing the 
generating monomial of the polar code viewed as a monomial code.
For this purpose denote by $W$ the memoryless channel for which the polar code is devised. Its input alphabet is binary
and its output alphabet is denoted by $\Yc$ and for the sake of simplifying a little bit the discussion, 
it is also assumed to be 
discrete. We assume that the channel is symmetric meaning that 
 there exists a permutation $\pi$ of $\Yc$ which is also an involution ($\pi^{-1}=\pi$) and 
 $W(y|1) = W(\pi(y)|0)$ for all $y \in \Yc$.
 We define the Ar{\i}kan channel transforms $\plus{W}$ and $\moins{W}$ of $W$ which 
 are both binary-input memoryless symmetric channel with transitions probabilities specified by
 \begin{eqnarray*}
 \plus{W}(y_1,y_2,u_2|u_1) & \eqdef & \frac{1}{2} W(y_1|u_1) W(y_2|u_1 \oplus u_2) \\
 \moins{W}(y_1,y_2|u_2) & \eqdef & \frac{1}{2} \sum_{u_1 \in \F_2} W(y_1|u_1) W(y_2|u_1 \oplus u_2)
 \end{eqnarray*}
Here the output alphabet of $\moins{W}$ is $\Yc \times \Yc$ whereas the output alphabet 
of $\plus{W}$ is  $\Yc \times \Yc \times \F_2$.
We extend this definition to $W^{\uv}$ where $\uv$ is a sequence of $+$ and $-$ in a natural way by
 writing for instance 
 $$
 W^{+-} \eqdef \moins{\left(\plus{W}\right)}.
 $$
 We will overload this notation 
 by also denoting by $\W{g}{m}$ for a monomial $g$ in $\Mo$ the channel $\W{u_{m-1} \cdots u_0}{m}$ where $u_i = -$ if
 $g$ involves $x_i$ and $u_i= +$ otherwise.  
 For instance when $m=5$ then
 $W_5^{x_3x_1x_0}$ means $W^{+-+--}$.
 Finally we will also need to define the Bhattacharyya parameter $\Bha{W}$ of a binary-input symmetric channel $W$. It is given by
 $$
 \Bha{W} \eqdef \sum_{y \in \Yc} \sqrt{W(y|0) W(y|1)}
 $$
 With these definitions we can construct a polar code of length $n=2^m$ and dimension $k$ devised for a binary-input symmetric channel $W$. 
 \begin{definition}
 The polar code of length $n=2^m$ and dimension $k$ devised for the channel $W$ is the monomial code $\CC(I)$ 
 where $I$ is the set of $k$ monomials in $\Mo$ which take the $k$ smallest values $\Bha{W^g}$ among all 
 $g$ in $\Mo$.
 \end{definition}
 Note  that the output alphabet size of the channels $W^g$ is exponential in $m$ which makes this ranking rather delicate. 
 However there are efficient methods for computing these $k$ ``best'' channels, see for instance \cite{TV13} where 
 ranking is performed for the error probability which is arguably even more complicated to track than the 
 Bhattacharyya parameter.

\section{Decreasing Monomial Codes}  \label{sec:decreasing}
Polar codes and Reed-Muller codes are both monomial codes but this family is too large to explain
the intriguing algebraic properties of polar codes (for instance their very large automorphism group).
We also want to capture simple properties that give some insight about which monomials to choose
in a polar code and this for any channel. Of course, $\plus{W}$ is a much better channel than $\moins{W}$ and it 
is straightforward to prove based on this intuition that a polar code of nonzero dimension always involves the monomial
$1$ in its definition and this for every channel. We will prove guided by this ``principle'' that 
if $f$ divides $g$ and if $g$ is a monomial in the defining monomial set $I$ of the polar code $\CC(I)$ 
then $f$ also belongs to $I$. If we define the partial order between monomials induced by 
divisibility considerations, that is $f \weako g$ iff $f$ divides $g$, then the defining monomial set $I$ of a polar code
$C(I)$ is {\em decreasing}, meaning that if $g \in I$ any $f$ such that $f \weako g$ also belongs to $I$. 
Here the ``w'' in $\weako$ stands for ``weak'' (as in weak order) to distinguish between this divisibility partial order and a
much finer order that we will introduce below. 
This divisibility order can already be used to prove that the permutation group of a polar code contains a group isomorphic to $\Z_2^m$ for a polar
code of length $2^m$. This proves that polar codes admit a $1$-transitive permutation group for instance. But we can go much beyond this by introducing a much finer ordering of the monomials  than the
divisibility ordering $\weako$. The order we will consider is the following

\begin{definition}
Two  monomials of the same degree are ordered as 
$x_{i_1}\dots x_{i_s}  \preceq x_{j_1}\dots x_{j_s}$
if and only if  for any 
$\ell \in \{1,\dots,s\}$, it holds $i_\ell \le j_\ell$
where we assume $i_1 < \dots < i_s$ and $j_1 < \dots < j_s$.


This partial order is extended to monomials of different degrees
through divisibility, namely $f \preceq g$ if and only if there is a
divisor $g^*$ of $g$ such that $f \preceq g^*\weako g$.
\end{definition}
\medskip
From this definition,  for any monomial $f$ of $\Mon$ the constant polynomial $1$ satisfies the inequality $1  \preceq f$.  We also have that $x_0\preceq x_1 \preceq \dots \preceq x_{m-1}.$ The \emph{interval} $[f , h]$ where $f$ and $h$ are in $\Mon$ with
$f\preceq h$ is the set of monomials $g \in \Mon$ such that $f \preceq g
\preceq h$. We will also need the following definition
\begin{definition}
  A set $I \subseteq \Mon$ is \emph{decreasing}  if and only if  ($f \in I$ and $g \preceq f $) implies
  $g \in I$.   
  A set $I \subseteq \Mon$ is {\em weakly decreasing} if and only if ($f \in I$ and $g \weako f $) implies
  $g \in I$.   

\medskip
  
  When $I \subseteq \Mon$ is a \emph{decreasing} set then $\CC(I)$ is called \emph{decreasing monomial code}.
  It is called a {\em weakly decreasing monomial code} if $I$ is weakly decreasing.
\end{definition}
We provide some useful facts about this order.
\begin{proposition}
Let $f$, $g$ in $\Mo$ such that $f\preceq g$ and let us set $f^*$, $g^*$ as the monomials in $\Mo$ 
such that $f = f^* \gcd(f,g)$ and $g = g^* \gcd(f,g)$, then we have:
\[
f^* \preceq h^*.
\]
Furthermore, for any $h$ in $\Mo$ such that $\gcd(f,h) = \gcd(g,h) = 1 $, we have:
\[
f h \preceq g h.
\]
\end{proposition}

\begin{remark}
The condition on $h$ to be prime with $f$ and $g$ is essential. 
Indeed, for instance $x_0 \preceq x_1$ but
$ x_0 x_1 \not \preceq  x_1^2$ since $x_1^2 = x_1$.
\end{remark}

Reed-Muller codes are decreasing codes as it is shown in what follows. 
\begin{proposition}
\label{prop:RM_decreasing}
All Reed-Muller codes are decreasing monomial codes given by
\begin{equation}
\label{eq:RM_decreasing}
 \RM(r,m)=\CC\left([1, x_{m-r}\cdots{} x_{m-1}]\right).
\end{equation}.
\end{proposition}

\begin{IEEEproof}
The set $[1, x_{m-r}\cdots x_{m-1}]$ is clearly decreasing and we are left to prove 
\eqref{eq:RM_decreasing}. Let $I$ be the set of monomials in $\Rm$ of degree at most $r$.
We have $\RM(r,m)=\CC(I)$. Note now that 
$x_{m-r}\dots x_{m-1}$ belongs to $I$ and that all monomials $f$ of degree at most $r$ are smaller than or equal to
this monomial 
$$
f \preceq x_{m-r}\cdots{} x_{m-1}.
$$
This implies 
$$
I \subseteq [1, x_{m-r} \cdots{} x_{m-1}]
$$
Moreover no monomial of degree greater than $r$ can be smaller than $x_{m-r}\dots x_{m-1}$. Therefore we have
$$
I = [1, x_{m-r}\cdots{} x_{m-1}].
$$
\end{IEEEproof}

It will turn out that polar codes devised for any binary-input symmetric channel are decreasing monomial codes.
\begin{theorem}\label{th:polar_decreasing}
Polar codes  are decreasing monomial codes.
\end{theorem}

We will start by proving a weaker statement whose ingredients and lemmas used in the proof will be essential
for the proof of Theroem~\ref{th:polar_decreasing}.
The weaker statement is the following proposition.
\begin{proposition}\label{prop:polar_weakly_decreasing}
Polar codes are weakly decreasing monomial codes.
\end{proposition}

\subsection{Proof of Proposition \ref{prop:polar_weakly_decreasing}}

The proof of this proposition relies on the notion of concatenated and degraded channel that we now recall.
\begin{definition}[Concatenated channels]
Let  $W:\Xc \rightarrow \Yc$
and $W':\Yc \rightarrow \Zc$ be two memoryless channels such that the input alphabet of $W'$  is equal
to the output alphabet of $W$. The concatenation of $W$ with $W'$ is denoted by 
$W' \circ W$ and is a memoryless channel $W":\Xc \rightarrow \Zc$ with transition probabilities specified by
$$
W"(z|x) = \sum_{y \in \Yc} W'(z|y)W(y|x).
$$
\end{definition}
\begin{definition}[Degraded channel] 
Let $W:\Xc \rightarrow \Yc$
and $W':\Xc \rightarrow \Zc$ be two memoryless channels, both with input alphabet 
$\Xc$ and respective output alphabets $\Yc$ and $\Zc$. 
We say that $W'$ is a channel degradation of $W$, if and only if there exists a memoryless
channel $W":\Yc \rightarrow \Zc$ such that $W'=W"\circ W$, that is
$$
W'(z|x) = \sum_{y \in \Yc} W"(z|y)W(y|x).
$$
We write
$$
W' \degrad W
$$
to denote that $W'$ is degraded with respect to $W$.
\end{definition}

It is easily shown that ``degraded'' is a transitive relation
\begin{equation}\label{eq:transitive}
\left. \begin{array}{lcr} W & \degrad & W' \\ W' & \degrad & W" \end{array} \right\} \Rightarrow W \degrad W".
\end{equation}
The following lemmas will also be very helpful
\begin{lemma}[{\cite[p207]{RU08}}]
\label{lem:Bha_degradation}
If $W \degrad W'$ then $\Bha{W'} \leq \Bha{W}$.
\end{lemma}

\begin{lemma}[{\cite[Lemma 4.7]{K09},\cite[Lemma 5]{TV13}}]
\label{lem:plus_moins_degradation}
Let $W$ be some binary input channel  and let $W'$ be a degradation of this channel ($W' \degrad W$).
In such a case
\begin{eqnarray*}
\moins{W'} & \degrad & \moins{W}\\
\plus{W'} & \degrad & \plus{W}.
\end{eqnarray*}
\end{lemma}

From this lemma we easily obtain that
\begin{corollary}
\label{cor:f_degradation}
Let $W$ be some binary input channel $W$ and let $W'$ be a degradation of this channel: $W' \degrad W$.
For any monomial $f$ in $\Mo$ we have
$$
\Wg{W'}{f}{m} \degrad \W{f}{m}.
$$ 
\end{corollary}
The crucial observation for the proof of Proposition \ref{prop:polar_weakly_decreasing} is that
\begin{lemma}\label{lem:moins_degraded}
For any binary-input symmetric channel $W:\{0,1\} \rightarrow \Yc$ we have
$$
\moins{W} \degrad W \degrad \plus{W}.
$$
\end{lemma}
\begin{remark}

\begin{itemize}
\item[]
\item The property $W \degrad \plus{W}$ is well known. It is mentioned for instance in \cite[p.268]{S11g} and proved in 
\cite[Lemma 4]{G15}).
\item Lemma \ref{lem:moins_degraded} has been proved in \cite[Lemma 4]{G15} when $W$ is an erasure channel. This follows
directly from the fact that in this case all three channels $\moins{W},W,\plus{W}$ are 
erasure channels with decreasing erasure probabilities and that an erasure channel of error probability $p$ is 
a degraded version of an erasure channel of probability $p'\leq p$.
It is used in \cite{G15} to prove that the index sets of polar codes devised for the erasure channel are quasi-similar 
(see \cite[Proposition 7.1]{G15}). For more general channels quasi-similarity is only proved in \cite{G15} to the ``right half''
(see \cite[Proposition 7]{G15}). Note that Lemma \ref{lem:moins_degraded} can be used together with the proof technique of Proposition 7.1 in \cite{G15} to strengthen this Proposition 7.1 
namely to show quasi-similarity for all binary input symmetric channels.  
\end{itemize}
\end{remark}
\begin{IEEEproof}
First of all let us recall why we obviously have $W \degrad \plus{W}$. Consider the output $(y_1,y_2,x)$ of the $\plus{W}$ channel 
when a bit $b$ has been sent throught it. By definition $y_1$ is the result of sending $b$ through the $W$-channel. Therefore if we define $W'$ as the channel 
$W': \Yc \times \Yc \times \{0,1\}\rightarrow \Yc$ which takes $(y_1,y_2,b)$, erases $y_2$ and $b$ 
to send just $y_1$, we clearly have that 
$W = W'\circ \plus{W}$.

Let us prove now that $\moins{W} \degrad W$ by constructing a channel $W''$ such that $W''\circ W = \moins{W}$.
The channel $W''$ is defined as follows. It takes as input  $y_1$.
At that point a bit $x$ is drawn uniformly at random and sent through channel $W$ to get some value $y_2$.
Then $W''$ outputs the pair
$(y_1,y_2)$ when $x=0$ and $(\pi(y_1),y_2)$ when $x=1$. Here $\pi$ is the involution acting on $\Yc$ such that
$W(y|1) = W(\pi(y)|0)$.
Figure \ref{fig:Wsec} summarizes the channel $W''\circ W$ derived from it. 
\begin{figure}[h!]
\centering
\caption{A figure summarizing $W''\circ W$ \label{fig:Wsec}}
\includegraphics[height=4cm]{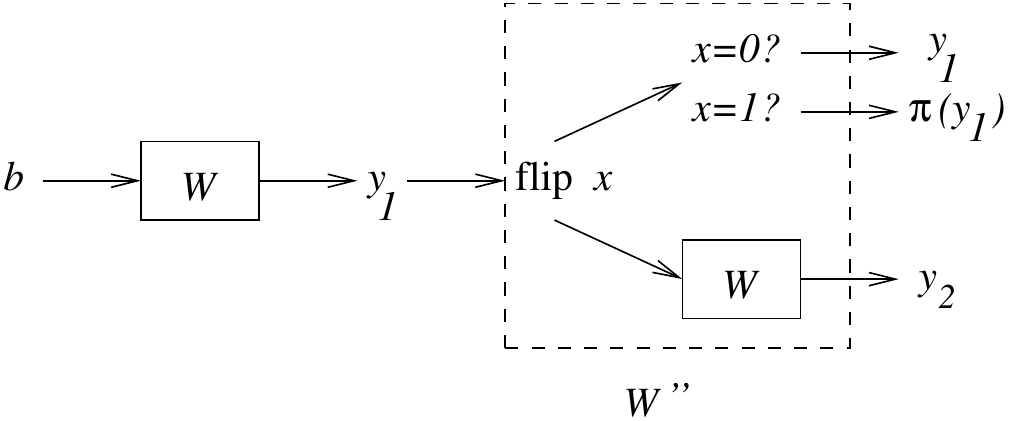} 
\end{figure}

Now sending $b$ through channel $W''\circ W$ and receiving $y_1 y_2$ happens in two cases
\begin{itemize}
\item
when $x=0$, this happens when $y_1$ has been received after sending $b$ through $W$ and $y_1,y_2$ has been received after sending $y_1$ through $W"$. This means that for $W"$, $y_2$ has been received
with the second use of $W$ when $x$ has been sent through it;
\item when $x=1$, this happens when $\pi(y_1)$ has been received after sending $b$ through $W$ (so that $W "$ changes 
$\pi(y_1)$ into  $\pi(\pi(y_1))=y_1$) and $y_2$ has been received
with the second use of $W$ when $x$ has been sent through it.
\end{itemize}
This implies that
\begin{eqnarray*}
(W''\circ W)(y_1,y_2|b) & = & \prob(x=0) \prob(\text{receiving $(y_1,y_2)|b$ was sent, $x=0$}) + \\
& &  \prob(x=1) \prob(\text{receiving $(y_1,y_2)|b$ was sent, $x=1$})\\
& = & \prob(x=0)W(y_1|b)W(y_2|x=0)+\prob(x=1)W(\pi(y_1)|b)W(y_2|x=1)  \\
& = & \frac{1}{2} \left\{W(y_1|b)W(y_2|x=0)+W(\pi(y_1)|b)W(y_2|x=1)\right\}  \\
& = & \frac{1}{2} \left\{W(y_1|b)W(y_2|x=0)+W(y_1|1 \oplus b)W(y_2|x=1)\right\} \text{ (by symmetry of $W$)} \\
& = & \frac{1}{2} \sum_{u \in \F_2} W(y_1|u)W(y_2|u \oplus b)\\
& = & \moins{W}(y_1,y_2|b)
\end{eqnarray*}
This computation shows that $W'' \circ W$ is precisely the channel $\moins{W}$.
\end{IEEEproof}

From this lemma we deduce the following result from which Proposition \ref{prop:polar_weakly_decreasing} is
essentially a corollary.
\begin{proposition}
\label{prop:decreasing_degrading} Let $W$ be a binary input symmetric channel. Let $f$ and $g$ be two monomials of $\Rm$. If
$f \weako g$ then
$$
\W{g}{m} \degrad \W{f}{m}.
$$
\end{proposition}
\begin{IEEEproof}
This follows by induction on $m$. When $m=1$ we just have to prove that 
\begin{equation}
\label{eq:m_1}
\W{x_0}{1} \degrad \W{1}{1}.
\end{equation}
Note that
\begin{eqnarray*}
\W{x_0}{1} & = & \moins{W}\\
\W{1}{1} & = & \plus{W}
\end{eqnarray*}
The inequality \eqref{eq:m_1} follows 
directly from Lemma \ref{lem:moins_degraded}.
Assume now that Proposition \ref{prop:decreasing_degrading} holds for some positive integer $m$.
Let $f$ and $g$ be in $\Mop$ and assume that $f \weako g$. Let  us define the following monomials:
\begin{eqnarray*}
f_{0 \dots m-1}  &= & \gcd \left(f,\prod_{i=0}^{m-1} x_i \right)
\\
g_{0 \dots m-1}  &=  & \gcd \left(g,\prod_{i=0}^{m-1} x_i \right)
\end{eqnarray*}

\begin{eqnarray*}
f_{m+1} & = & \left \{
			\begin{array}{ll}
				x_0 & \text{if } x_{m+1}  \text{ divides } f, \\
				1  & \text{otherwise.}
			\end{array}
			\right.
\\
g_{m+1} & = & \left \{
			\begin{array}{ll}
				x_0  &\text{if } x_{m+1}  \text{ divides } g, \\
				1      & \text{otherwise.}
			\end{array}
			\right.
\end{eqnarray*}
Note now that
\begin{eqnarray*}
\W{f}{m} & = & \Wg{\Wb{f_{m+1}}{1}}{f_{0\dots m-1}}{m}\\
\W{g}{m} & = & \Wg{\Wb{g_{m+1}}{1}}{g_{0\dots m-1}}{m}
\end{eqnarray*}
Since $f_{0\dots m-1}$ divides $g_{0\dots m-1}$ we have by the induction hypothesis
\begin{equation}
\label{eq:induction}
 \Wg{\Wb{f_{m+1}}{1}}{g_{0\dots m-1}}{m} \degrad  \Wg{\Wb{f_{m+1}}{1}}{f_{0\dots m-1}}{m}
\end{equation}
Since $f_{m+1} \weako g_{m+1}$ we have
$$
\W{g_{m+1}}{1} \degrad \W{f_{m+1}}{1}.
$$
From Corollary \ref{cor:f_degradation} we deduce that
\begin{equation}
\label{eq:induction2}
\Wg{\Wb{g_{m+1}}{1}}{g_{0\dots m-1}}{m} \degrad  \Wg{\Wb{f_{m+1}}{1}}{g_{0\dots m-1}}{m}
\end{equation}
From \eqref{eq:induction} and \eqref{eq:induction2} we deduce that 
$$
 \Wg{\Wb{g_{m+1}}{1}}{g_{0\dots m-1}}{m} \degrad  \Wg{\Wb{f_{m+1}}{1}}{f_{0\dots m-1}}{m}
$$
which proves the statement of the proposition for $m+1$.
\end{IEEEproof}

We are ready now to finish the proof of Proposition \ref{prop:polar_weakly_decreasing}.
\begin{IEEEproof}[Proof of Proposition \ref{prop:polar_weakly_decreasing}]
Let $\CC(I)$ be a polar code generated by the set of monomials $I$ devised for a channel $W$.
Let $f$ and $g$ be two monomials such that $g$ is in $I$ and $f \weako g$.
From Proposition \ref{prop:decreasing_degrading} we know that 
$$
\W{g}{m} \degrad \W{f}{m}.
$$
By applying Lemma \ref{lem:Bha_degradation} we deduce that 
$$
\Bha{\W{f}{m}} \leq \Bha{\W{g}{m}}.
$$
This implies that $f$ also belongs to $I$.
\end{IEEEproof}

\subsection{Proof of Theorem \ref{th:polar_decreasing}}

In order to prove Therorem~\ref{th:polar_decreasing} we will need to have a finer understanding 
of the $\W{f}{m}$'s. This is achieved through the following proposition which can be proved easily by induction
and which is basically given in \cite{A09} (we have just transformed a little bit the usual definition of the 
channel with our polynomial formalism).

\begin{lemma}\label{lem:channel_model}
The model of channel $\W{f}{m}$ can be described as follows. 
To send a bit $x$, one first chooses a binary word $\av =(a_g)_{g \in \Mm}$ of length $2^m$  indexed by the monomials of $\Mo$    uniformly at random in $\{0,1\}^{2^m}$.
Let $f= x_{i_1} \cdots{} x_{i_t}$. Let $\av'=(a'_g)_{g \in \Mo}$ be the binary word of length $2^m$ with 
$a'_g = a_g$ with the exception of $a_f$ where $a'_f = x$.
One computes $\bv \eqdef \av' \Gm_m$ and sends the bits of $\bv$ through channel $W$ to obtain 
a vector $\yv \in \Yc^{2^m}$.
Consider the total order $>$ on the monomials induced by the order on the integers and by viewing a
monomial $g = x_{j_1} \dots x_{j_t}$ as the integer $2^{j_1} + \dots + 2^{j_t}$. The corresponding output of the 
channel $\W{f}{m}$ is then $\yv$ together with $(a_g)_{g : g>f}$.
\end{lemma}

Theorem \ref{th:polar_decreasing} heavily relies on Proposition \ref{prop:decreasing_degrading} on one 
hand and the following lemma on the other hand.
\begin{lemma}\label{lem:plus_moins_moins_plus_degrading}
Let $W$ be a symmetric binary input channel. In such a case for any positive integer $m$ we have
$$
\W{x_1 x_2 \dots x_{m-2} x_{m-1}}{m} \degrad \W{x_0x_1\dots x_{m-2}}{m}.
$$
\end{lemma}
\begin{IEEEproof}
We use Lemma \ref{lem:channel_model} and consider $\W{x_0x_1\dots x_{m-2}}{m}$. As explained in this lemma, for sending a
bit $c$ this channel
 consists in:
\begin{enumerate}
\item choosing a binary word $\av =(a_g)_{g \in \Mm}$ of length $2^m$      uniformly at random in $\{0,1\}^{2^m}$;
\item replacing the entry $a_{x_0x_1\dots x_{m-2}}$ by $c$ to obtain a binary word $\av'$;
\item computing $\bv \eqdef \av' \Gm_m$;
\item sending the entries through channel $W$ to obtain a vector $\yv=(y_g)_{g \in \Mo} \in \Yc^{2^m}$;
\item outputting $\yv$ and $(a_g)_{g : g>{x_0x_1\dots x_{m-2}}}$.
\end{enumerate}
Let $\tau : \Mo  \longrightarrow  \Mo$ be a permutation on the set of monomials in $\Mo$ defined by
\begin{eqnarray*}
\tau(x_{i_1} \cdots x_{i_t}) & = & x_{(i_1-1)\pmod{m}} \cdots x_{(i_t-1)\pmod{m}}.
\end{eqnarray*}
It is straightforward to check that if we concatenate this channel with a channel $W'$ which consists in reordering 
$\yv$ as $\yv^\tau \eqdef (y_{\tau(g)})_{g \in \Mm}$  and erasing all the entries of $(a_g)_{g : g>{x_0x_1\dots x_{m-2}}}$ with the exception 
of $a_{x_0 \dots x_{m-1}}$ we obtain the channel $\W{x_1 x_2 \dots x_{m-2} x_{m-1}}{m}$.
\end{IEEEproof}

Before giving a slightly more general statement, let us introduce some notation which will be very helpful.
 \begin{notation}
Let $f=x_{i_1} \dots x_{i_s}$ be a monomial in $\Mo$. We denote by 
$f_{[a..b]}$ the monomial $\Pi_{i_j : a \leq i_j \leq b} x_{i_j}$ and by 
$f^t$ the monomial $x_{i_1+t} \dots x_{i_s+t}$.
\end{notation}

With this notation at hand we have
\begin{lemma}
Let $f$ and $g$ be two monomials of $\Mo$ of the same degree such that 
(i) $f \preceq g$ (ii) $f_{[0..i-1]}=g_{[0..i-1]}$, $f_{[i+t+1..m-1]}=g_{[i+t+1..m-1]}$, 
(iv) $f_{[i..i+t]} =  x_{i}x_{i+1} \dots x_{i+t-1}$, (v) $g_{[i..i+t]} =  x_{i+1}x_{i+2} \dots x_{i+t}$.
Then 
$$
\W{g}{m} \degrad \W{f}{m}
$$
 More generally in the case of two monomials $f$ and $g$ satisfying (ii) and (iii) and
 \begin{equation}
 \label{eq:hyp}
 \W{g_{[i..i+t]}^{-i}}{t+1} \degrad  \W{f_{[i..i+t]}^{-i}}{t+1}
 \end{equation}
 then 
 $$
 \W{g}{m} \degrad \W{f}{m}.
 $$
\end{lemma}

\begin{IEEEproof}
We can write in the first case
\begin{eqnarray*}
\W{f}{m} & = & \Wg{\Wgb{\Wb{f_{[i+t+1..m-1]}}{m-i-t-1}}{x_0 x_{1} \dots x_{t-1}}{t+1}}{f_{[0..i-1]}}{i}\\
\W{g}{m} & = & \Wg{\Wgb{\Wb{f_{[i+t+1..m-1]}}{m-i-t-1}}{x_1 x_{2} \dots x_{t}}{t+1}}{f_{[0..i-1]}}{i}.
\end{eqnarray*}
We can apply Lemma \ref{lem:plus_moins_moins_plus_degrading} to $\W{f_{[i+t+1..m-1]}}{m-i-t-1}$ to deduce that
\begin{equation*}
\Wgb{\Wb{f_{[i+t+1..m-1]}}{m-i-t-1}}{x_{1} x_{2} \dots x_{t}}{t+1}
\degrad
\Wgb{\Wb{f_{[i+t+1..m-1]}}{m-i-t-1}}{x_0 x_{1} \dots x_{t-1}}{t+1}.
\end{equation*}
By applying Corollary \ref{cor:f_degradation} we obtain
$$
\Wg{\Wgb{\Wb{f_{[i+t+1..m-1]}}{m-i-t-1}}{x_1 x_{2} \dots x_{t}}{t+1}}{f_{[0..i-1]}}{i}
\degrad
\Wg{\Wgb{\Wb{f_{[i+t+1..m-1]}}{m-i-t-1}}{x_0 x_{1} \dots x_{t-1}}{t+1}}{f_{[0..i-1]}}{i}.
$$
The second statement follows by a similar reasoning but uses now \eqref{eq:hyp} directly instead of 
using Lemma \ref{lem:plus_moins_moins_plus_degrading}.
\end{IEEEproof}
This lemma can now be used to prove by induction on $m$ that
\begin{lemma}
\label{lem:almost_order}
Let $W$ be a symmetric binary input channel. Let $m$ be a positive integer and let $f$ and $g$ be two monomials of $\Mo$
that are of same degree and such that $f \preceq g$. Then 
$$
\W{g}{m} \degrad \W{f}{m}.
$$
The second statement of the lemma follows from a similar reasoning.
\end{lemma}

\begin{IEEEproof}
When $m=1$ the two monomials are necessarily equal and we are done.
Assume now that the property we want to prove holds for all positive integers up to some positive integer $m$.
Consider now two monomials $f$ and $g$ of $\Mop$ with the same degree $t$ and such that 
$f \preceq g$.
We can write
\begin{eqnarray*}
f & = & x_{i_1} \dots x_{i_t} \\
g & = & x_{j_1} \dots x_{j_t}
\end{eqnarray*}
with $i_1 < i_2 < \dots < i_t$ and $j_1 < j_2 < \dots j_t$.
Let $i_l$ be the smallest index among $\{i_1,\dots,i_t\}$ such that
$i_l < j_l$. If such an index does not exist we have $f=g$ and therefore
$\W{g}{m+1} \degrad \W{f}{m+1}$ which is what we want to prove.
Let $i_s$ be the largest index greater than or equal to $i_l$ such that
$$
i_j = i_l + (j-l)
$$
for all $j$ in $\{l,l+1,\dots,s\}$.
In other words in such a case
$$
x_{i_l} \dots x_{i_s} = x_{i_l} x_{i_l+1} \dots x_{i_l + (s-l)}.
$$
Observe that we can write $f$ as
$$
f = f_{[0..i_l-1]} x_{i_l} x_{i_l+1} \dots x_{i_l + (s-l)} f_{[i_l+(s-l)+2..m]}.
$$
We can apply the previous lemma and obtain
\begin{equation}
\label{eq:first_part}
\W{f_{[0..i_l-1]} x_{i_l+1} x_{i_l+2} \dots x_{i_l + (s-l)+1}f_{[i_l+(s-l)+2..m]}}{m+1} \degrad 
\W{f}{m+1}.
\end{equation}
Observe now that $g$ is such that  
$$ 
 x_{i_l+1} x_{i_l+2} \dots x_{i_l + (s-l)+1} f_{[i_l+(s-l)+2..m]} \preceq g_{[i_l..m]}.
$$
This comes from the fact that all the $j_u$'s for $u \in \{l,l+1,\dots,s\}$  necessarily satisfy
$j_u \geq i_u+1$, since this is true for $u=l$ and can be shown for values that are larger by noting that
 $j_u \geq j_l+(l-u) \geq i_l+1+(l-u) = i_u +1$.
 We can apply the induction hypothesis to the pair
 $g_{[i_l..m]}^{-i_l}$ and $x_{1} \dots x_{(s-l)+1} f_{[i_l+(s-l)+2..m]}^{-i_l}$ since
 $$
  x_1 \dots x_{(s-l)+1} f_{[i_l+(s-l)+2..m]}^{-i_l} \preceq g_{[i_l..m]}^{-i_l}  
 $$
 and therefore
 $$
\W{g_{[i_l..m]}^{-i_l}}{m-i_l+1}  \degrad \W{x_1 \dots x_{(s-l)+1} f_{[i_l+(s-l)+2..m]}^{-i_l}}{m-i_l+1}.
 $$
 By applying Corollary \ref{cor:f_degradation} 
 we deduce that 
 \begin{equation}
 \label{eq:second_part}
 \W{g}{m+1}  \degrad \W{g_{[0..i_l-1]}x_{i_l+1} \dots x_{i_l+(s-l)+1} f_{[i_l+(s-l)+2..m]}}{m+1}.
 \end{equation}
 Using the fact that by definition $g_{[0..i_l-1]}=f_{[0..i_l-1]}$ and putting \eqref{eq:first_part} and
 \eqref{eq:second_part} together by using the transitivity of $\degrad$ we get
 $$
 \W{g}{m+1} \degrad \W{f}{m+1}
 $$
 which proves that the induction hypothesis also holds up to $m+1$.
 \end{IEEEproof}

We are ready now for the proof of Theorem \ref{th:polar_decreasing}.
\begin{IEEEproof}[{Proof of Theorem \ref{th:polar_decreasing}}]
Let $\CC(I)$ be a polar code generated by the set of monomials $I$ devised for a channel $W$.
Let $f$ and $g$ be two monomials such that (i) $g$ is in $I$ (ii) $f \preceq g$.
Assume first that $f$ and $g$ have the same degree.
In such a case we can apply Lemma \ref{lem:almost_order} and deduce that
$$
\W{g}{m} \degrad \W{f}{m}.
$$
When $f$ and $g$ are not of the same degree, we know that there exists a divisor $g^*$ of $g$ such that 
$g^*$ and $f$ have the same degree and $f \preceq g^*$.
By applying Lemma \ref{lem:almost_order} to the pair $(f,g^*)$ we deduce that
$$
\W{g^*}{m} \degrad \W{f}{m}.
$$
Since $g^*$ divides $g$ we know from Proposition \ref{prop:polar_weakly_decreasing} that
$$
\W{g}{m} \degrad \W{g^*}{m}.
$$
By transitivity of $\degrad$ we deduce again that 
$$
\W{g}{m} \degrad \W{f}{m}.
$$
Therefore in all cases we can apply Lemma \ref{lem:Bha_degradation} and obtain that 
$$
\Bha{\W{f}{m}} \leq \Bha{\W{g}{m}}.
$$
This implies that $f$ also belongs to $I$.
\end{IEEEproof}


\section{Structural Properties of Decreasing Monomial Codes}\label{sec:properties}

The algebraic formalism introduced in previous sections permits to reveal several interesting properties about decreasing monomial codes. In this section, we focus only on three important aspects: 
characterising the dual code, estimating the minimum distance and identifying a large subgroup of the permutation group.

\subsection{Duality}


It is readily seen that the dual of a monomial code is a polynomial code, but it is not necessarily a monomial code.
However the dual of a decreasing monomial code turns out to be a decreasing monomial code. 
In order to describe precisely the duality we will define some notation. The set of indices of 
the variables appearing in a  monomial $g \in \Mo$  is denoted by $\indices(g)$. Hence, we have
$$
g = \prod\limits_{i\in \indices(g)}x_i.
$$
%
%
The \emph{multiplicative complement} of a monomial $g \in \Mo$ denoted by $\check{g}$, 
is defined as: 
$$
\check{g} = \prod\limits_{i \in \{0,\dots,m-1\}\setminus \indices(g)}  x_i
$$ 
By extension for any subset $I \subseteq \Mo$, the set $\check{I} \subseteq \Mo$ denotes 
$\{ \check{f}~:~ f \in I \}$.

\begin{lemma} \label{lem:barineq}
For all $f$ and $g$ in $\Mo$, $f \preceq g$ if and only if  $\check{f} \succeq \check{g}$. 
\end{lemma}
%
%
\begin{IEEEproof}
Let $f$ and $g$ be two monomials in $\Mo$ such that $f\preceq g$. By definition, there exists a divisor 
$g^*$ in $\Mo$ of $g$ such that $\deg g^* = \deg f$  and $f \preceq g^*$. Let us set  $a = \dfrac{g}{g^*}$.
By observing that  $\check{f} \succeq \check{g^*}$ and 
$\check{g} = \gcd(\check{g^*}, \check{a})$, which means in particular $\check{g}$ divides $\check{g*}$, 
we can deduce by the triangle inequality that $\check{f} \succeq \check{g}$. 
\end{IEEEproof}

\begin{corollary}\label{cor:decreasing}
Let $I \subseteq \Mo$ be a decreasing set then  $\Mo \setminus \check{I}$ is a decreasing set. 
\end{corollary}
\begin{IEEEproof}
  Let $h$ be a monomial that belongs to $\Mo \setminus \check{I}$,
  and let $g\in\Mo$ be a monomial such that $g \preceq h$.  Assume by
  contradiction that $g \notin \Mo \setminus \check{I}$,
  i.e. $g\in\check{I}$. Then there exists $f \in I$ such that $g =
  \check{f}\preceq h$, which implies that $\check{h} \preceq f$ by
  Lemma ~\ref{lem:barineq}. Since $I$ is a decreasing set, $\check{h}
  \in I$, that is to say, $\check{{\check{h}}} = h \in \check{I}$
  which contradicts the assumption. Therefore $\Mo \setminus
  \check{I}$ is a decreasing set
\end{IEEEproof}

\begin{proposition}\label{prop:dual_monomial}  Let $\CC(I)$ be a decreasing monomial code, then its dual is a decreasing monomial code given by
$$
 \CC(I)^{\bot}=\CC( \Mo \setminus \check{I}).
 $$
\end{proposition}

\begin{IEEEproof}
As $\dim(\check{I})=\dim(I)$, 
  we have $\dim\CC(\Mo \setminus \check{I}) = \card{\Mo \setminus
    \check{I}} = N - \dim \CC(I)=\dim\CC(I)^{\bot}$,
     so we need to prove only one inclusion.

 Let $f  \in \Mo \setminus \check{I}$ and consider $g \in I$. Assume that $f g = x_0 \cdots{} x_{m-1}$. This means there exists $h \in \Mo$ such that $f = h \check{g}$, or equivalently $ \check{g}\preceq f$, that is to say $f \in \check{I}$ because $I$ is a decreasing set  (and thanks to Lemma~\ref{lem:barineq}). Hence  the inclusion
$\CC( \Mo \setminus \check{I}) \subseteq  \CC(I)^{\bot}$ is proved.
\end{IEEEproof}

Notice that this proposition yields the well known result about the dual of a Reed-Muller code
 \begin{eqnarray*}
 \RM(r,m)^\bot &=& \CC(\Mo\setminus[x_0 \dots x_{m-r-1};x_0\dots x_{m-1}]) \\
 &=&
 \CC\left([1; x_{r+1}\dots x_{m-1}]\right)=\RM(m-r-1,m).
 \end{eqnarray*}
A straightforward consequence of this is  that under some conditions, 
any decreasing monomial code is weakly self-dual. 
\begin{corollary}\label{cor:dual} 
Let $\CC(I)$ be a decreasing monomial code with $\card{I} \leq \frac{1}{2} 2^m$. Then 
$\CC(I) \subseteq \CC(I)^\bot$
if and only if  for any 
$f \in I, \;   \check{f}  \not \in I$.

\end{corollary}

Polar codes of rate (sufficiently) smaller than $1/2$ generally satisfy this assumption and in the 
case of rate greater than $\frac{1}{2}$ it is the dual of the polar code that satisfies this assumption.
This can be explained by looking at the polarization process that is used to choose the monomials 
defining the polar code. 

\subsection{Minimum Distance of Decreasing Monomial Codes}

The estimation of the minimum distance of a decreasing monomial code will need the following notion.

\begin{definition}Let $\CC(I)$ be a decreasing monomial code over $m$ variables.
We let 
\begin{eqnarray*}
\rmo(\CC(I)) & \eqdef & \max \big \{r ~|~ \RM(r,m) \subseteq \CC(I) \big \}\\
\rpl(\CC(I)) & \eqdef & \min \big \{r ~|~  \CC(I) \subseteq \RM(r,m) \big \}
\end{eqnarray*}
\end{definition}

It is readily checked that another way of defining these quantities is that $\rmo$ is the largest $r$ for which the monomial
$x_{m-r}\cdots x_{m-1}$ is in $I$. On the other hand  $\rpl$ is the largest integer $r$ for 
which $x_0 \cdots x_{r-1}$ is in $I$.
These quantities are related to the minimum distance of a decreasing monomial code and its dual through the following result
\begin{proposition}\label{pr:dmin}
Let $\CC(I)$ be a decreasing monomial code over $m$ variables. We have the following properties:
\begin{enumerate}
\item
The minimum distance of $\CC(I)$ is equal to
$2^{m-\rpl(\CC(I))}$. 
\item $\rmo(\CC(I)^\perp)$ and $\rpl(\CC(I)^\perp)$ satisfy the equalities:
\begin{eqnarray*}
\rmo(\CC(I)^\perp) & = & m-1- \rpl(\CC(I))\\
\rpl(\CC(I)^\perp) & = & m-1- \rmo(\CC(I))
\end{eqnarray*}
\item The minimum distance of $\CC(I)^\perp$ is equal to 
$2^{\rmo(\CC(I))+1}$
\end{enumerate}
\end{proposition}

\subsection{Permutation Group} 

Applying an affine permutation to a monomial code yields a polynomial code but not necessarily a 
monomial code. Furthermore, polynomial codes and monomial codes may have a trivial permutation group.
However by considering the subclass
 of decreasing monomial codes we obtain codes with a very large permutation group which is
the \emph{lower triangular affine group}. Before giving its precise definition, we 
introduce some notation. Binary square matrices with $m$ rows (and $m$ columns) are denoted by $\F_2^{m \times m}$.  Let us recall that  a bijective affine transformation over $\F_2^m$ can be represented by a pair $(\Am,\bv)$ where $\Am$ lies in the general linear group $\GL(m,2)$ and $\bv$ in $\F_2^m$. The action of $(\Am,\bv)$ on a monomial $g$ is denoted by $(\Am,\bv) \cdot{} g$. It basically consists in replacing each monomial $x_i$ by a ``new" monomial  $y_i$ defined by:
\[
y_i = x_i + \sum_{j=0}^{m-1} a_{ij} x_j + b_i.
\]
In the case of decreasing monomial codes, we are interested in a subclass of these transformations that are  \emph{lower triangular}. We recall that a matrix $\Am = (a_{i,j}) $ is lower triangular if
$a_{i,j}  = 0$ whenever $j > i$.

\begin{definition}
The set of affine transformations over $\F_2^m$ of the 
form $\xv \mapsto \Am \xv + \bv$ where $\Am \in \F_2^{m \times m}$ is a lower triangular binary matrix 
with $a_{i,i} = 1$ and 
$\bv \in \F_2^m$ forms a group called the \emph{lower triangular affine group} $\Alow$.
\end{definition}

\begin{theorem}\label{th:permutation-group}
The permutation group of a decreasing monomial code in $m$ variables contains $\Alow$.
\end{theorem}
\begin{IEEEproof}
Let $\CC(I)$ be a decreasing monomial code and let $(\Am,\bv)$
be in $\Alow$. The action of $(\Am,\bv)$ where $\Am = (a_{i,j}) \in \F_2^{m\times m}$ and $\bv \in \F_2^m$ can be viewed as a change of variables
where $x_i$ is replaced by the variable $y_i$ defined by
\[
y_i = x_i + \sum_{j=0}^{i-1} a_{ij} x_j + b_i.
\]
Hence if $x_{i_1} \cdots{} x_{i_s}$ belongs to $I$  with $0 \leq i_1 < \cdots{} < i_s \leq m-1$ 
then $y_{i_1} \cdots{} y_{i_s}$ is a linear combination involving only monomials 
of the form $\prod_{i \in J} x_i$ where $J$ describes the powerset of $\{i_1,\dots{},i_s\}$. In particular,
$\prod_{i \in J} x_i$ is in $I$ since $I$ is decreasing and therefore
$\eval(y_{i_1} \cdots{} y_{i_s})$ belongs to $\CC(I)$, which terminates the proof.

\end{IEEEproof}

\begin{remark}
Although the permutation group of a Reed-Muller code is well-known, 
the question remains open for decreasing monomial codes. 
\end{remark}


\section{Minimum Weight Codewords}\label{sec:MinWords}
\subsection{Structure of orbits under the action of $\Alow$}

A natural object when dealing with group actions is the orbit of an
element. 
We denote by
\[\orbit{g} = \{(\Am,\bv)\cdot g | \, (\Am,\bv)\in\Alow\} \text{ for } g \in \Mon\]
the orbit of a monomial $g$ under the action of $\Alow$.When $g$ is equal to the monomial $x_i$ then its orbits is
of the form $\left\{x_i+\sum\limits_{j=0}^{i-1} a_{j}x_j+b\; |\; a_j\text{ and } b\in \F_2\right\}$.
A consequence is that the cardinality of the orbit of $x_i$ equals $2^{i+1}.$ 

When the degree of $g$ is greater than 1 counting the number of elements in the orbit is less obvious. The reason why the task is more complicated comes from the fact that the stabilizer subgroup of $\Alow$ with respect to $g$ is not trivial. The following example illustrates this fact.

\begin{exemple}
Let $g=x_0x_1$ then by definition
$\Alow . g = \left\{ (x_0 + b_0)(x_1 + a_{1,0}x_0 + b_1) | b_0,a_{1,0},b_1\in\F_2\right\}$.
We remark that there are two group elements  in $\Rm$ that leave $g$ invariant:
$(x_0 + b_0)(x_1 + a_{1,0}x_0 + b_1) = x_0x_1 $
if and only if $b_0=0$ and $a_{1,0}=b_1$, in other words
$x_0x_1=x_0(x_1+x_0+1).$ In this case there are 4 distincts polynomials in the
orbit of $x_0x_1$ which are $x_0 x_1$, $x_0(x_1+1)$, $(x_0+1)x_1$ and  $(x_0+1)(x_1+1)$. 
\end{exemple} 

\begin{definition} For any $g$ from $\Mo$ we define $\Alow_g$ as the subgroup
 of $(\Am,\bv) \in \Alow$  such that:
\[
 b_i = 0 \text{ if } i \not \in \indices(g) 
 ~~~\text{ and }~~ 
 a_{ij} = 
 \left \{ 
 	\begin{array}{l}
		0 \text{ if } i \not \in  \indices(g) \\
		0 \text{ if }  j \in \indices(g).
	 \end{array}
 \right.
\] 
\end{definition}

\begin{proposition}\label{prop:lta_quotien} For any monomial $g$ in $\Mon$  the orbit of $g$ under the action of $\Alow$
is equal to the orbit of $g$ under the action of $\Alow_g$:
\begin{equation} \label{eq:orbit}
\Alow \cdot g = \Alow_g\cdot g.
\end{equation}
\end{proposition}


\begin{IEEEproof} 
The inclusion $\Alow_g\subseteq \Alow$ is clear.  We prove the
  converse inclusion by induction on $\deg g$. For $\deg g =0$ and
  $1$, it is clear that \eqref{eq:orbit}  holds. Let us assume 
  that is true  for any monomial of degree $ \leq d$ where $d \ge 1$. Let $g$ be a monomial in 
  $\Mon$ of degree $d +1$.  Let us consider $(\Am,\bv)$ in $\Alow$ and for any $i$ in $\indices(g)$
  we define $y_i$ as:
  $$
  y_i = x_i + \sum_{t = 0}^{i-1} a_{i t} x_t + b_i
    $$
    Hence we have $(\Am,\bv)\cdot g = \prod\limits_{j \in \indices{(g)}} y_j$. 
    Now let $i$ be the \emph{maximum} element of $\indices(g)$. We then have:
     $$
  	y_i =  x_i + \sum_{t \in \indices(g)}a_{i t} x_t  + \sum_{t = 0,  t \notin \indices(g)}^{i-1} a_{i t} x_t 
	+ b_i
     $$
Using the fact that $f^2 = f$ for any $f$ in $\Rm$, we also have:
\begin{IEEEeqnarray*}{rCl}
 	\prod_{j \in \indices{(g)}} y_j & = &  y_i \prod_{t \in \indices{(g)}, t \neq i} y_t \\
  	& = &   \left( x_i + \sum_{t \in \indices(g)} a_{i t} (x_t + 1 + y_t) \right.\\
	&  & \; \; \left. {+} \sum_{t = 0,  t \notin \indices(g)}^{i-1} a_{i t} x_t  + b_i \right) \prod_{t \in \indices{(g)}, t \neq i} y_t
 \end{IEEEeqnarray*}
Since $\prod_{t \in \indices{(g)}, t \neq i} y_t $ is of degree $d$ then by induction assumption, 
 there exists $(\Am^*,\bv^*) \in \Alow_g$ such that for any $t \in \indices(g)$ and $t \neq i$, it holds:
 $$
  y_t = x_t + \sum_{s = 0, s\notin \indices(g)}^{t-1} a^*_{t s} x_s + b^*_t.
  $$
 This implies in particular that we can write:
 $$
 \sum_{t \in \indices(g)} a_{i t} (x_t + 1 + y_t)  = \sum_{t \in \indices(g)} \sum_{s = 0, s\notin \indices(g)}^{t-1} a_{i t}  (a^*_{t s} x_s + b^*_t + 1) 
 $$
 This last equation only involves variables $x_s$ with $0 \leq s < i$ and $s \notin \indices(g)$. 
 Hence we can find a binary vector $(a^*_{i 0},\dots{},a^*_{i,i-1})$ with $a_{i t} = 0$ if $t \in \indices(g)$, and $b_i^* \in \F_2$ such that:
 $$
 \prod_{t \in \indices{(g)}} y_t= \prod_{t \in \indices{(g)}} 
\left(  x_t + \sum_{s = 0, s\notin \indices(g)}^{t-1} a^*_{t s} x_s + b^*_t \right).
 $$
 This last equality proves $\Alow\cdot g \subseteq \Alow_g\cdot g$ and concludes the proof.
 \end{IEEEproof}

\begin{proposition}\label{prop:unicity_orbit}
For any $g\in\Mo$ we have 
$$\left|\orbit{g}\right|=\left|\Alow_g\right|.$$
\end{proposition}

\begin{IEEEproof} 
From Proposition \ref{prop:lta_quotien} we have that  $\card{\orbit{g}}\le \card{\Alow_g\cdot g}.$

Let $g \in \Mon$ and let us consider  $(\Am,\bv)$ and $(\Am',\bv')$ in $\Alow_g$. We prove that if
$(\Am,\bv)\cdot g = (\Am',\bv')\cdot g$ in $\Rm$ then  $\Am=\Am'$ and

This comes from the fact that in the polynomial
  $(\Am,\bv)\cdot g\in\Rm$, the coefficient of
  $x_j\prod\limits_{k\in\indices{(g)}, k\neq i}x_k$ when $i\in\indices{(g)}$ and
  $j\notin\indices{(g)}$  is exactly $a_{ij}$ and
  the coefficient of $\prod\limits_{k\in\indices{(g)}, k\neq i}x_k$ is
  $b_{i}$. This proves that $(\Am,\bv)=(\Am',\bv')$.

Therefore there is a bijection between the two sets and counting the number of elements in the orbit of $g$ is equivalent to counting the number of pairs $\Am,\bv\in \Alow_g.$
\end{IEEEproof}

In order to give the cardinality of an orbit we use a well-known combinatorial object called the Ferrers diagram (or Young diagram).  

\begin{definition}
A \emph{Young diagram} is a finite collection of boxes arranged in left-justified rows, with the rows sizes weakly increasing. 
\end{definition}
    
\begin{figure}[H]
\begin{center}
\caption{Young diagrams inside a $2\times 3$ grid \label{fig:young23}}
\end{center}
\begin{center}
\begin{tikzpicture}[inner sep=0in,outer sep=0in]
\node(m)
{\begin{varwidth}{5cm}{
\begin{ytableau}
 ~&&\\
~ &&\\
 \end{ytableau}}\end{varwidth}};
\end{tikzpicture}
\quad
\begin{tikzpicture}[inner sep=0in,outer sep=0in]
\node(m)
{\begin{varwidth}{5cm}{
\begin{ytableau}
 ~&&\\
*(blue! 40) * &&\\
 \end{ytableau}}\end{varwidth}};
\end{tikzpicture}
\quad
\begin{tikzpicture}[inner sep=0in,outer sep=0in]
\node(m)
{\begin{varwidth}{5cm}{
\begin{ytableau}
 *(blue! 40) *&&  \\
*(blue! 40) * & & \\
 \end{ytableau}}\end{varwidth}};
\end{tikzpicture}
\quad
\begin{tikzpicture}[inner sep=0in,outer sep=0in]
\node(m)
{\begin{varwidth}{5cm}{
\begin{ytableau}
 ~&&  \\
*(blue! 40) * &*(blue!40)* & \\
 \end{ytableau}}\end{varwidth}};
\end{tikzpicture}
\end{center}
\begin{center}
\begin{tikzpicture}[inner sep=0in,outer sep=0in]
\node(m)
{\begin{varwidth}{5cm}{
\begin{ytableau}
 *(blue! 40) *&&  \\
*(blue! 40) * &*(blue!40)* & \\
 \end{ytableau}}\end{varwidth}};
\end{tikzpicture}
\quad
\begin{tikzpicture}[inner sep=0in,outer sep=0in]
\node(m)
{\begin{varwidth}{5cm}{
\begin{ytableau}
 *(blue! 40) *&*(blue! 40) *&  \\
*(blue! 40) * &*(blue! 40) * & \\
 \end{ytableau}}\end{varwidth}};
\end{tikzpicture}
\quad
\begin{tikzpicture}[inner sep=0in,outer sep=0in]
\node(m)
{\begin{varwidth}{5cm}{
\begin{ytableau}
~&&  \\
*(blue! 40) * & *(blue! 40)*&*(blue! 40) * \\
 \end{ytableau}}\end{varwidth}};
\end{tikzpicture}
\quad
\begin{tikzpicture}[inner sep=0in,outer sep=0in]
\node(m)
{\begin{varwidth}{5cm}{
\begin{ytableau}
 *(blue! 40) *&&  \\
*(blue! 40) * & *(blue! 40)*&*(blue! 40) * \\
 \end{ytableau}}\end{varwidth}};
\end{tikzpicture}
\end{center}
\begin{center}
\begin{tikzpicture}[inner sep=0in,outer sep=0in]
\node(m)
{\begin{varwidth}{5cm}{
\begin{ytableau}
 *(blue! 40) *&*(blue! 40) *&  \\
*(blue! 40) * &*(blue! 40) * &*(blue! 40) * \\
 \end{ytableau}}\end{varwidth}};
\end{tikzpicture}
\quad
\begin{tikzpicture}[inner sep=0in,outer sep=0in]
\node(m)
{\begin{varwidth}{5cm}{
\begin{ytableau}
 *(blue! 40) *&*(blue! 40) *&*(blue! 40) *  \\
*(blue! 40) * &*(blue! 40) * &*(blue! 40) * \\
 \end{ytableau}}\end{varwidth}};
\end{tikzpicture}
\end{center}
\end{figure}
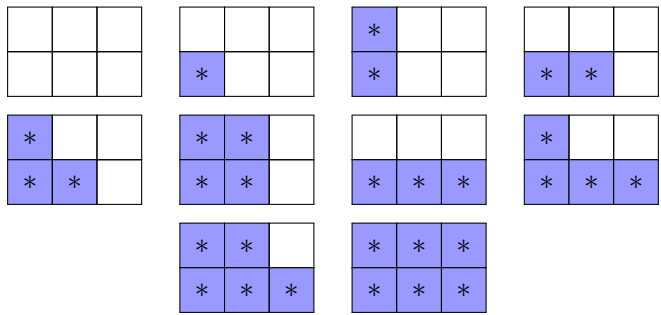

\medskip

The diagram is generally used to represent a partition $\lambda$ of
 integers. In the above figure we have the following partitions
from left to right: $\epsilon, (1), (1,1), (2), (1,2), (2,2), (3),
(1,3), (2,3), (3,3).$ 


We construct a bijection between Young diagrams in grids of size
$d\times (m-d)$ and monomials of degree $d$ in $m$ variables.
More precisely if $(\Am,\bv)\in \Alow_g $, then by definition of $\Am$ the rows
$i\notin\indices{(g)}$ and the columns $j\in\indices{(g)}$ contains
only a 1 on the diagonal (and 0 elsewhere). If we remove from $\Am$ the
rows $i\notin\indices{(g)}$ and the columns $j\in\indices{(g)}$, we
get a $d\times (m-d)$ matrix with possible non-zero coefficients
exactly inside the boxes of the associated Ferrers diagram.
\begin{proposition} 
  For any integers $m, d$ with $1\le d \le m$, there is a bijection
  between monomials in $\Mon$ of degree $d$ and Young diagrams inside
  the $d\times (m-d)$ grid.
\end{proposition}
\begin{IEEEproof}
   If $g=x_{i_1}\cdots x_{i_d}\in\Mon$ is a monomial of
  degree $d$, then the partition associated to $g$ is $\lambda_g =
  (i_d-(d-1), i_{d-1}-(d-2),\dots, i_1-0)$ inside the $d\times (m-d)$
  grid. It is a partition since $i_k - (k-1) \ge i_{k-1} - (k-2)$.

  If $\lambda =(\lambda_d,\dots,\lambda_1)$ is a partition inside the
  $d\times (m-d)$ grid with $\lambda_d\ge \lambda_{d-1}\ge \dots \ge
  \lambda_1\ge 0$, then the monomial $g$ associated to it is
  $g=x_{i_1}\cdots x_{i_d}$ where $i_k=\lambda_k +
  (k-1)>\lambda_{k-1}-1+(k-1)=i_{k-1}$.
\end{IEEEproof}

\begin{exemple}

 Let $m=5$, $g=x_1x_4$ then the partition associated to $g$ is 
 $\lambda_g=(4-1,1-0)=(3,1)$ and it's Young diagram in the $2\times 3$ grid is 
    \begin{center}
    \begin{tikzpicture}[inner sep=0in,outer sep=5in]
      \node(m) {\begin{varwidth}{5cm}{
\begin{ytableau}
 *(blue! 40) *  & &\\
*(blue! 40) *  &*(blue! 40) *   &*(blue! 40) *  \\
\end{ytableau}}\end{varwidth}};
\end{tikzpicture}.
 \end{center}
We illustrate the same bijection using  the matrix\[\Am=\begin{pmatrix}1&0&0
    &0 &0\\a_{10}&1&0 &0 &0\\0&0 &1&0&0 \\0&0&0 &1 &0\\ a_{40}&0&a_{42}&a_{43}
    &1\end{pmatrix}.\] After deleting the rows corresponding to $x_0,
  x_2, x_3$ and the columns corresponding to $x_1, x_4$, we get
  $\begin{pmatrix}a_{10} &0 &0\\a_{40} &a_{42}&a_{43} \end{pmatrix}$ which corresponding to
  the $8^{th}$ Ferrers diagram from Figure \ref{fig:young23}.
 We deduce that there are $2^4$ different matrices
$\Am$ in $\Alow_g$, and $2^2$ different vectors $\bv$ which
gives $|\orbit{x_1x_4}|=2^6$.

On the other hand, the monomials associated to the Young diagrams from
Figure~\ref{fig:young23} are (in the same order):
$x_0x_1$, 
$x_0x_2$, $x_1x_2$, 
$x_0x_3$, 
$x_1x_3$, 
$x_2x_3$, 
$x_0x_4$, 
$x_1x_4$, $x_2x_4$ and $x_3x_4$.
\end{exemple}

We denote by $\lambda_g$ the Ferrers diagram corresponding to $g$ and $\left|\lambda_g\right|$ the size of a diagram, that is to say the number of * in the diagram. Thanks to Proposition~\ref{prop:unicity_orbit} we can state the following.
\begin{proposition}\label{pr:orbti_mon_young}
The cardinality of the orbit of $g$ under the action of $\Alow$ is
\[\left|\Alow \cdot g\right|=2^{\ \deg(g)+\left|\lambda_g \right|}\]
\end{proposition}

\subsection{The minimum weight codewords of a decreasing monomial code.} 

Characterizing the minimum weight codewords is often quite difficult and there are few families of codes where the structure of the minimum weight codewords is well known. 
%
In the case of decreasing monomial codes the subgroup $\Alow$ gives enough information to understand the structure of the minimum weight codewords.  
We suppose that $\CC(I)$ is a decreasing monomial code and we denote by $I_{\rpl}=\{f\in I\; |\; \deg(f)=\rpl\}$ the set of monomials in $I$ of
maximal degree. From
Proposition~\ref{pr:dmin}, the set of minimum weight codewords
is $$\Wm=\{\cv \in \CC(I) ~ | ~ |\cv|=2^{m-\rpl}\}.  $$


\begin{proposition}\label{pr:min_codewords}
\[\Wm= \left \{\eval(P) ~|~  \exists f \in I_\rpl, \; P\in \Alow \cdot f   \right\}. \]
\end{proposition}
\begin{IEEEproof}
The $\supseteq$ inclusion comes from the fact that $\Alow$ acts on $\CC(I)$ as a permutation, and for any $f\in I_\rpl$, $\eval(f)$ has weight $2^{m-\rpl}$.

As for the $\subseteq$ inclusion, consider an arbitrary element
$\eval(P)\in\Wm$. From~\cite{KTA76} we know that an element of weight $2^{m-\rpl}$ is the evaluation of a product of $\rpl$ independent linear forms  $P = \prod\limits_{j=1}^\rpl \ell_j.$ Each linear form can be written as 
\[
\ell_j=x_{i_j}+\sum\limits_{0\le k <i_j}a_{j k}x_{k}+\epsilon_{j}.
\] 
If all the maximal variables $x_{i_j}$ in the $\rpl$ linear forms are pairwise distinct then $P\in \orbit{(x_{i_1}\dots x_{i_\rpl})}.$ 

Now suppose that this assumption is not true and consider without loss of generality two linear forms having the same maximum variable $\ell_{1}=x_{i_2}+\sum\limits_{k<i_2}a_{1 k}x_{k}+\epsilon_{1}$ and $\ell_{2}=x_{i_2}+\sum\limits_{k<i_2}a_{2 k}x_{k}+\epsilon_{2}.$ Let $\ell^*_{1}=x_{i_2}+\ell_{1}$ and $\ell^*_{2}=x_{i_2}+\ell_{2}.$ It is clear that $\ell_{1}^*+\ell_{2}^*\not = 0$ because of the independence condition. Using the relation $f^2=f$ for any $f\in\Rm$ we can rewrite
\begin{align*}
\ell_{1}\ell_{2}&=\left(\ell_{1}+\ell_{2}+1\right)\ell_{2}=\left(\ell_{1}^*+\ell_{2}^*+1\right)\ell_{2}= \ell^{'}_{1}\ell_{2}
\end{align*}   
where the maximum variable of $\ell^{'}_{1}$ is strictly smaller than
$x_{i_2}$ and the two forms are independent.

By induction we can prove that any product of $\rpl$ linear forms can be rewritten in a product of $\rpl$ linear independant forms with distincts maximal variable.
\end{IEEEproof}

Furthermore we prove that for any two distinct monomials the intersection of their orbits is the empty set and conclude with the formula counting the minimum weight codewords of a decreasing monomial code.  

\begin{lemma}\label{lem:intersect_orbit}
Let $f$ and $g$ be two different monomials in $\Mon.$  Then the intersection of their orbits is equal to the empty set.
\[\Alow\cdot  f\cap \Alow \cdot g =\emptyset.\]
\end{lemma}
\begin{IEEEproof}
If the two monomial can be compared with respect to our order then we can consider without loss of generality that $f\preceq g.$ Using the definition of $\Alow$ we have that any polynomial in the orbit of $g$ contains the monomial $g$ and any polynomial is the orbit of $f$ does not contain the monomial $g.$ So it is impossible to find a polynomial belonging to both orbits. 

If the two monomials can not compared the proof works in the same way.   
\end{IEEEproof}

\begin{theorem}
Let $\CC(I)$ be a decreasing monomial code, 
then the number of minimum weight codewords in $\CC(I)$ equals
\[\left|\Wm\right|=2^{\rpl}\sum\limits_{g\in I_\rpl}2^{\left|\lambda_g \right|}.\]
\end{theorem} 
\begin{IEEEproof}

Use Proposition \ref{pr:min_codewords} and Lemma \ref{lem:intersect_orbit}
\end{IEEEproof}

\begin{corollary}
The number of minimum weight codewords in  $\RM(r,m)$ equals 
\[W_{min}(\RM(r,m))=2^r\binom{m}{r}_2\]

where $\binom{m}{r}_2=\dfrac{(2^m-1)\dots(2^m-2^{r-1})}{(2^r-1)\dots(2^r-2^{r-1})}$ is the Gaussian binomial coefficient.

\end{corollary}

\begin{IEEEproof}
  Recall that $\binom{m}{r}_2$ represents the number of
  $r$-dimensional subspaces of $\F_2^m.$ The problem of counting the
  number of $r$-dimensional subspaces of $\F_2^m$ is equivalent to
  counting the number of $r\times m$ matrices of rank $r$ in reduced
  echelon form. Each matrix gives rise to a Young diagram inside the
  $r\times(m-r)$ grid and each diagram $\lambda$ can be obtained from
  $2^{\left|\lambda\right|}$ matrices. So we have the following
  combinatorial identity:
\begin{equation}
\binom{m}{r}_2=\sum\limits_{\lambda \subset r\times(m-r)}2^{\left|\lambda \right|}.
\end{equation} 

Moreover we recall that $\RM(r,m)=\CC\left([1, x_{m-r}\dots x_{m-1}]\right),$ which implies that the Young diagrams corresponding to all the maximum degree monomials of a Reed-Muller Code are all the possible diagrams in the $r\times(m-r)$ grid. Thus we have 
 \[\Wm(\RM(r,m))=2^{r} \sum\limits_{\lambda\subset r\times(m-r) }2^{\left|\lambda \right|}=2^r\binom{m}{r}_2\]  
 
\end{IEEEproof}
%
%


\section{Acknowledgement}
We thank Bernhard Geiger for pointing out to us his work \cite{G15} on the fractality 
of polar and Reed-Muller codes.

\bibliographystyle{IEEEtran}
\bibliography{codecrypto}

\end{document}